\documentclass[12pt]{iopart}
\usepackage{graphicx}

\def\be{\begin{equation}}
\def\ee{\end{equation}}

\begin{document}

\title[Gaussian factor in the distribution
arising from the nonextensive statistics]{Gaussian factor in the
distribution arising from the nonextensive statistics approach to
fully developed turbulence}
 \author{A~K~Aringazin\footnote[3]{Also at Department of Mechanics and
Mathematics, Kazakhstan Division, Moscow State University, Moscow
119899, Russia.} and M~I~Mazhitov}
 \address{Department of Theoretical Physics, Institute for
Basic Research, Eurasian National University, Astana 473021
Kazakhstan} \eads{\mailto{aringazin@mail.kz}, \mailto{mmi@emu.kz}}

\begin{abstract}
We propose a simple modification, the Gaussian truncation, of the
probability density function which was obtained by Beck (2001) to
fit the experimental distribution of fluid particle acceleration
component from fully developed fluid turbulence inspired by the
framework of nonextensive statistical mechanics with the
underlying gamma distribution of the model parameter. The modified
distribution is given a phenomenological treatment and provides a
good fit to new experimental results on the acceleration of fluid
particle reported by Crawford, Mordant, and Bodenschatz. We
compare the results with the new fit which was recently obtained
by Beck (2002) with the use of the underlying log-normal
distribution of the model parameter.
\end{abstract}

 \pacs{05.20.Jj, 47.27.Jv}

\maketitle

%% 1
\section{Introduction}
\label{Sec:Introduction}

Recently, C. Beck \cite{Beck} have studied application of Tsallis
nonextensive statistics formalism~\cite{Tsallis}  to turbulent
flows and achieved a good agreement with the high-precision
Lagrangian data on the component of acceleration measured by
Porta, Voth, Crawford, Alexander, and
Bodenschatz~\cite{Bodenschatz}. Remarkably, no fit parameters have
been used by Beck to reproduce distribution of the acceleration
$a$ of a tracer particle advected by the turbulent
flow, while %%{\it ad hoc}
the stretch exponential fit provides an excellent agreement with
the experiment but requires three free parameters for this
purpose~\cite{Bodenschatz,Bodenschatz2},
\be\label{1}
P(a) = C \exp\left[-a^2/[(1
+\left|{b_1a}/{b_2}\right|^{b_3})b_2^2]\right],
\ee
where $b_1=0.513\pm 0.003$, $b_2= 0.563\pm 0.02$, and $b_3=
1.600\pm 0.003$ are fit parameters, and $C \simeq 0.733$ is a
normalization constant. The big $|a|$ asymptotics is given by
$P(a) \sim \exp[-|a|^{0.4}]$.

The main idea underlying the generalized statistical mechanics
approach to the turbulence is simply to introduce fluctuations of
the energy dissipation rate~\cite{Beck2,Wilk} which follow gamma
(chi-square) or some other appropriate distribution, into the
stochastic dynamical model framework.

This type of models belongs to the class of purely temporal
stochastic models of Lagrangian turbulence and corresponds to the
universality of the developed turbulence which is expected to
occur in the inertial range only statistically. Accordingly,
velocity and acceleration of an individual fluid particle are
treated as random variables, and one is interested in their time
evolution and probability distribution functions, or multipoint
correlation functions.  By the universality (Kolmogorov 1941,
Heisenberg 1948, Yaglom 1949), steady-state statistics of velocity
increments in the inertial range and statistics of acceleration do
not depend on details of the forcing and the dissipation (the
dependence is postulated only via mean energy injection rate).

In a physical context, we describe fluid particle dynamics in
statistically homogeneous and isotropic turbulence in terms of a
Brownian like motion. Such models are formulated in the Lagrangian
frame and naturally employ Langevin type equation and the
associated Fokker-Planck equation, which can be derived under
certain assumptions either in Stratonovich or Ito formulations.
Under the stationarity condition, a balance between the energy
injected at the integral scale and the energy dissipated by
viscous processes at the Kolmogorov scale, one can try to solve
the Fokker-Planck equation to find stationary one-point
probability density function of the variable. This function as
well as its moments can then be compared with the Lagrangian
experimental data and direct numerical simulations (DNS) of the
three-dimensional Navier-Stokes equation.

In contrast to the usual Wiener process, the components of
acceleration do not merely wander in a random way with a complete
self-similarity at all scales but were found to reveal a highly
intermittent behavior, which can be seen from a strongly
non-Gaussian character of the experimental probability density
function of acceleration component of the tracer
particle~\cite{Bodenschatz,Bodenschatz2}. This requires a
consideration of some specific Langevin type equation and
assigning some specific properties to additive and multiplicative
noises, which are used to model the force.

With the simple choice of white-in-time noises such models fall
into the class of Markovian models allowing well established
Fokker-Planck approximation while the consideration of finite-time
correlated noises and memory effects requires a deeper analysis.
Approximation of a short-time correlated noise by the zero-time
correlated one is usually made due to the time scale hierarchy
emerging due to physical analysis of the system and experimental
data.

In the Lagrangian framework some known examples of such models are
due to Sawford~\cite{Sawford}, Beck~\cite{Beck,Beck2,Beck4}, and
Reynolds~\cite{Reynolds}. In the Eulerian framework (fixed probe)
Langevin type equations for the velocity increments in space and
for logarithm of the stochastic energy dissipation rate have been
recently considered by Renner {\it et al.}~\cite{Friedrich}.

Specifically, the class of models considered in the present paper
is featured by the introduction of {\em random} intensity of the
additive noise and/or random drift coefficient (random intensity
of a multiplicative noise, in a more general consideration) that
was found to imply strongly non-Gaussian marginal distributions of
the acceleration which are associated to the Lagrangian
intermittency, a phenomenon that developed turbulent flows exhibit
when considering Lagrangian velocity increments at small
timescales. Effectively, this approach allows one to account for
longtime effects which are known to be important in describing the
Lagrangian intermittency~\cite{Mordant0206013}. The model
parameter $\beta>0$ appears here as the ratio between the drift
coefficient and intensity of the additive noise. The marginal
distribution $P(a)$ is obtained by integrating out $\beta$ in the
Gaussian distribution of the acceleration, $P(a|\beta)$, a
stationary solution of the Fokker-Planck equation, under the
assumption that $\beta$ follows some judiciously chosen
distribution.

In general, this approach assumes two characteristic timescales,
one associated to a short-time scale, of the order of Kolmogorov
time $\tau_\eta$, and the other associated to a longtime scale, of
the order of Lagrangian integral time $T_L$, in a way similar to
that assumed in the Sawford model~\cite{Sawford}. (The Sawford
model yields however independent Gaussian distributions for
Lagrangian accelerations and velocities with zero means and
constant variances.) The Fokker-Planck equation makes a link
between the stochastic dynamical treatment and statistical
approach. We refer the reader to
references~\cite{Beck,Beck2,Beck4} for details of this approach.

Earlier work on such type of models is due to Castaing, Gagne, and
Hopfinger~\cite{Castaing}, referred to as the Castaing model, in
which log-normal distribution of the variance of a Gaussian
distributed intermittent variable has been used to derive the
marginal distribution without reference to a stochastic dynamical
equation.

It should be stressed that the class of one-dimensional Langevin
toy models considered in the present paper suffers from the lack
of physical interpretation of coefficients and additive noise in
the context of the three-dimensional Navier-Stokes equation and
turbulence; see, e.g., reference~\cite{Kraichnan0305040}. It is
however quite a theoretical challenge to make a link between the
Navier-Stokes equation and such surrogate Langevin models, which
are, of course, far from being full statistical model of turbulent
dynamics of fluid particle. Strong and nonlocal character of
Lagrangian particle coupling due to pressure effects makes the
main obstacle to derive turbulence statistics from the
Navier-Stokes equation. Recent attempt in this direction is due to
the Batchelor-Proudman type stochastic distortion theory approach
to the Navier-Stokes equation by Laval, Dubrulle, and
Nazarenko~\cite{Laval} from which one can draw parallels with the
considered class of models.

In the present paper we suggest a simple modification of the
chi-square model, a Gaussian truncation of the resulting power law
distribution. We compare results of this model with those of the
chi-square model and recently suggested log-normal
model~\cite{Beck4}, and the experimental data on Lagrangian
acceleration statistics. Our consideration is restricted to a
stationary one-point distribution function. Two-point statistical
analysis is of much interested and can be made elsewhere.

Despite the considered models are not provided by a direct link to
the three-dimensional Navier-Stokes turbulence, they can be viewed
as prototypical stochastic models to interpret statistical data on
the recently measured tracer particle Lagrangian acceleration
data. Here we note only that due to the Navier-Stokes equation in
the Lagrangian frame the acceleration in the inertial range of
fully developed turbulence is driven by the pressure gradient term
which dominates the viscous term. This is confirmed by both the
agreement between the acceleration measurements and DNS results on
the pressure gradient, and recent verification of the longstanding
Heisenberg-Yaglom scaling of a component of acceleration, $\langle
a^2\rangle
%%\simeq a_0\epsilon^{3/2}\nu^{-1/2}
= a_0{\bar u}^{9/2}\nu^{-1/2}L^{-3/2}$, where ${\bar u}$ denotes
root mean square (rms) velocity, for the Taylor microscale
Reynolds numbers $R_\lambda>500$. Thus, the stochastic dynamical
modelling of the acceleration can be used to shed some light on
statistical properties of hydrodynamical forces acting on the
tracer particle.

The layout of the paper is as follows. In
Section~\ref{Sec:ChiSquare} we outline results of the chi-square
and the log-normal models. In Section~\ref{Sec:ChiSquareGaussian}
the chi-square Gaussian model is proposed, and a comparison with
the recent experimental data on acceleration statistics is made.
In Section~\ref{Sec:Conclusions} we summarize the obtained results
and make remarks.

%% 2
\section{The chi-square and log-normal models}
\label{Sec:ChiSquare}

Recently, Crawford, Mordant, and Bodenschatz \cite{Bodenschatz2}
reported new experimental results (for $R_\lambda=690$), which are
slightly different from that of the earlier
experiment~\cite{Bodenschatz} due to bigger amount of data points
collected, and pointed out that the Beck's marginal
distribution~\cite{Beck},
\be\label{2}
P(a) = \frac{C}{(1+\frac{1}{2}V_0(q-1)a^2)^{1/(q-1)}},
\ee
which is based on the gamma distribution of the fluctuating
dissipation energy rate, does not correctly capture the tails of
the experimental distribution function. Here, $q=3/2$ (Tsallis
entropic index), $V_0=4$, and $C=2/\pi$ is a normalization
constant, all the parameter values are due to the theory (not
fitted to the experimental curve). An essential discrepancy is
clearly seen from the contribution $a^4P(a)$ to the fourth order
moment.

In the recent paper, Beck \cite{Beck4} suggested to use log-normal
distribution instead of the gamma distribution, and the resulting
marginal distribution,
\be\label{3}
P(a) = \frac{1}{2\pi s}\int_{0}^{\infty}\! d\beta\
\beta^{-1/2}
\exp\left[-\frac{(\mathrm{ln}\frac{\beta}{m})^2}{2s^2}\right]
e^{-\frac{1}{2}\beta a^2},
\ee
where $m=\exp[s^2]$ provides a unit variance and $s$ is free
parameter, was found to be in a good agreement, for $s^2=3.0$,
with new experimental $P(a)$ and $a^4P(a)$ data
\cite{Bodenschatz2} and direct numerical simulations of the
Navier-Stokes equation ($R_\lambda=380$)~\cite{Gotoh}.
Particularly, much better fit has been achieved for the observed
low probability tails as compared to that of the chi-square model.
Provided that $\beta$ is related to the stochastic energy
dissipation rate, this model is in a correspondence with the
Kolmogorov 1962 theory, which assumes a log-normal distribution of
the stochastic energy dissipation rate.

In general, a log-normal distribution arises when many independent
random variables are combined in a multiplicative way. This
distribution is followed by an exponential of normally distributed
random variable, that is $\beta$ in equation~(\ref{3}) is treated
as an exponential of some normal variable. Therefore, from
equation~(26) of~\cite{Beck4} we see that
\be\label{ln}
\ln\beta=\ln[2C_0^{-2}a_0\nu^{1/2}\epsilon^{-3/2}],
\ee
obtained by a comparison with the Sawford model, is assumed to be
a normally distributed random variable with zero mean. Here, $a_0$
is the Kolmogorov constant which measures acceleration variance
($a_0\simeq 6$ due to Lagrangian experiments~\cite{Bodenschatz}),
$C_0$ is the Lagrangian structure constant ($C_0\simeq 7$ due to
the DNS data~\cite{Reynolds}), $\nu$ is the kinematic viscosity,
and $\epsilon$ is the energy dissipation rate per unit mass. The
argument of the above logarithm is then assumed to be a random
variable with nonzero mean.

The value $s^2=3.0$ used for the fitting can be understood as a
number of independent normal variables with unit variance and zero
mean which represents three space directions of the energy
dissipation rate~\cite{Beck4}. From this point of view, the value
of the free parameter, $s^2=3$, is predicted by theory and thus
there are no fit parameters in the marginal distribution
(\ref{3}).

The integral in equation~(\ref{3}) is calculated only numerically
that makes a statistical mechanics analysis less handful as
compared to the distribution (\ref{2}), which corresponds to the
well known Tsallis nonextensive distribution with the associated
Tsallis entropy. Nevertheless, Tsallis and Souza \cite{Tsallis2}
showed that the associated statistical mechanics can be
constructed for the log-normal distribution case as well.

In the next Section, we propose a simple modification of the
chi-square model (\ref{2}) which enables one to obtain better fit
to the very recent experimental data as compared with that of the
chi-square model.

%% 3
\section{The chi-square Gaussian model}
\label{Sec:ChiSquareGaussian}

In order to provide an analytically explicit anzatz, we suggest
the modified marginal distribution~\cite{Aringazin2},
\be\label{4}
P(a) =
\frac{C\exp[-a^2/a_c^2]}{(1+\frac{1}{2}V_0(q-1)a^2)^{1/(q-1)}},
\ee
which is obtained by a "Gaussian screening" of the marginal
distribution (\ref{2}). Again, $C$ is a normalization constant,
$V_0=4$, and $q=3/2$, while $a_c$ is a our free parameter, which
we use for a fitting.

From a phenomenological point of view, this modification is in
accord to the fact that Gaussian distributions are known to be
stationary solutions of the Sawford model~\cite{Sawford}, which
assumes some stochastic differential equation for acceleration
with non-fluctuating parameters. The Gaussian distribution is
known to be in agreement with the experimental distribution $P(a)$
for small accelerations $a$~\cite{Bodenschatz}.

An extension of the Sawford model to the case of fluctuating
parameters involved to Langevin type equation was recently made by
Reynolds~\cite{Reynolds}. A relationship between the class of
models (\ref{2}) and (\ref{3}), to which the model (\ref{4})
belongs, and Sawford model has been studied in the
work~\cite{Beck4}. A comparison of implications of the models
(\ref{2}), (\ref{3}) and the Reynolds model has been made in the
recent paper by Mordant, Crawford, and
Bodenschatz~\cite{Mordant0303003}.

Within the framework of Tsallis nonextensive statistics, the
parameter $q-1$ measures a variance of fluctuations. For $q\to 1$
(i.e., no fluctuations), equation~(\ref{4}) reduces to the
Gaussian distribution,
\be\label{gauss}
P(a) = C\exp[-(a_c^{-2}+\frac{1}{2}V_0)a^2],
\ee
while for $a_c\to \infty$ we obtain equation~(\ref{2}), so that
the proposed Gaussian truncated power law type distribution
(\ref{4}) makes a simple analytic interpolation between the two
results. This is the main phenomenological point to get a better
fit.

One can obtain the marginal distribution (\ref{4}) in a rigorous
way by using exactly the same approach as in the
reference~\cite{Beck4}. Namely, for a linear drift force, the
conditional probability density function
$P(a|\beta)=Z^{-1}(\beta)\exp[-\beta a^2/2]$ is replaced by
\be\label{5}
P(a|\beta)=Z^{-1}(\beta)\exp[-a^2/a_c^2 -\beta a^2/2],
\ee
to account for a non-fluctuating part, so that the averaging of
$P(a|\beta)$ over the $\chi^2$ distributed $\beta$ directly
implies the distribution (\ref{4}).

In essence, this means the redefinition, $\beta/2 \to
a_c^{-2}+\beta/2$, with some non-fluctuating part of the energy
dissipation rate being incorporated in the introduced parameter,
\be\label{6}
(a_c^{-2}+\frac{1}{2}\beta_0)^{-1} =
C_0^2a_0^{-1}\nu^{-1/2}\langle\epsilon\rangle^{3/2} =
C_0^2\frac{\langle a^2\rangle}{a_0^2},
\ee
where $\beta_0$ is an average of $\beta$ over the $\chi^2$
distribution, $V_0\propto\beta_0$. The r.h.s. of
equation~(\ref{6}) is due to the Heisenberg-Yaglom scaling
relation for a component of acceleration.

The above picture corresponds to a distribution of sum of squared
normal random variables, which have {\it nonzero} means
(noncentrality) and unit variances. From this point of view, the
relevance of the log-normal distribution~\cite{Beck4} can be
partially understood by the fact that it implicitly provides a
nonzero mean for the energy dissipation, equation~(\ref{ln}), that
meets experiments.

%% FIGURE 1
%%%%%%%%%%%%%%%%%%%%%%%%%%%%%%%%%%%%%%%%%%%%%%%%%%%%%%%
\begin{figure}[tbp!]
\begin{center}
\includegraphics[width=0.8\textwidth]{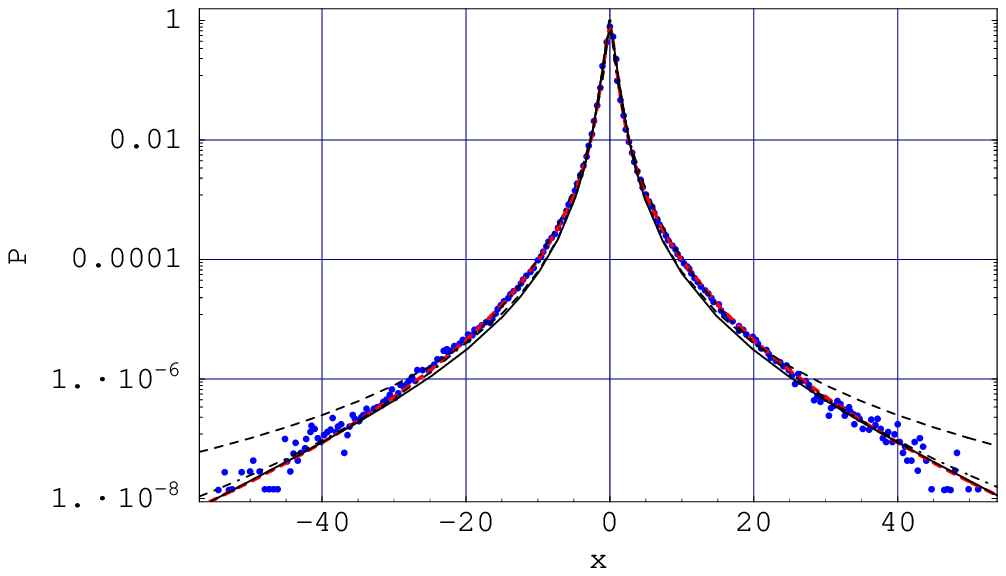}
\includegraphics[width=0.8\textwidth]{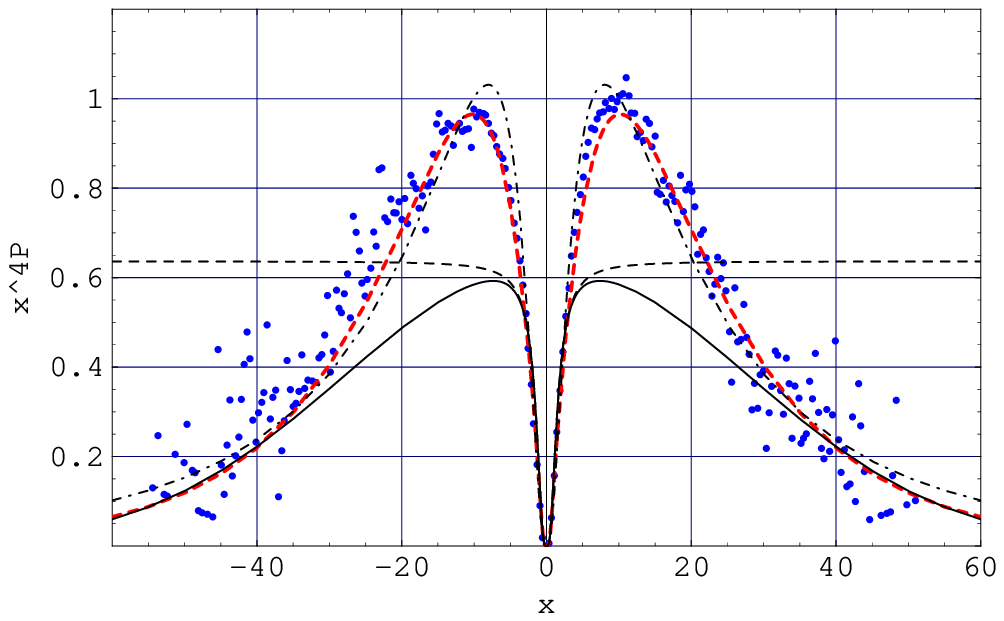}
\end{center}
\caption{ \label{Fig1} Top panel: lin-log plot of fluid particle
acceleration probability density functions, the stretched
exponential fit (\ref{1}) (thick dashed line), the chi-square
model (\ref{2}) (dashed line, $q=3/2$), the log-normal model
(\ref{3}) (dot-dashed line, $s^2=3$), and the chi-square Gaussian
model (\ref{4}) (solid line, $q=3/2, a_c=39.0$). Bottom panel: the
associated contributions to the fourth order moment, $a^4P(a)$.
Dots: the experimental data points for the transverse (with
respect to the large-scale symmetry axis of the studied flow)
component of acceleration measured by Crawford, Mordant, and
Bodenschatz~\cite{Bodenschatz2} ($R_\lambda=690$, the acceleration
is low-pass filtered at 0.23$\tau_\eta$ with Gaussian kernel).
$x=a/\langle a^2\rangle^{1/2}$ denotes normalized Lagrangian
acceleration.}
\end{figure}
%%%%%%%%%%%%%%%%%%%%%%%%%%%%%%%%%%%%%%%%%%%%%%%%%%%%%%%

%% FIGURE 2
%%%%%%%%%%%%%%%%%%%%%%%%%%%%%%%%%%%%%%%%%%%%%%%%%%%%%%%
\begin{figure}[tbp!]
\begin{center}
\includegraphics[width=0.8\textwidth]{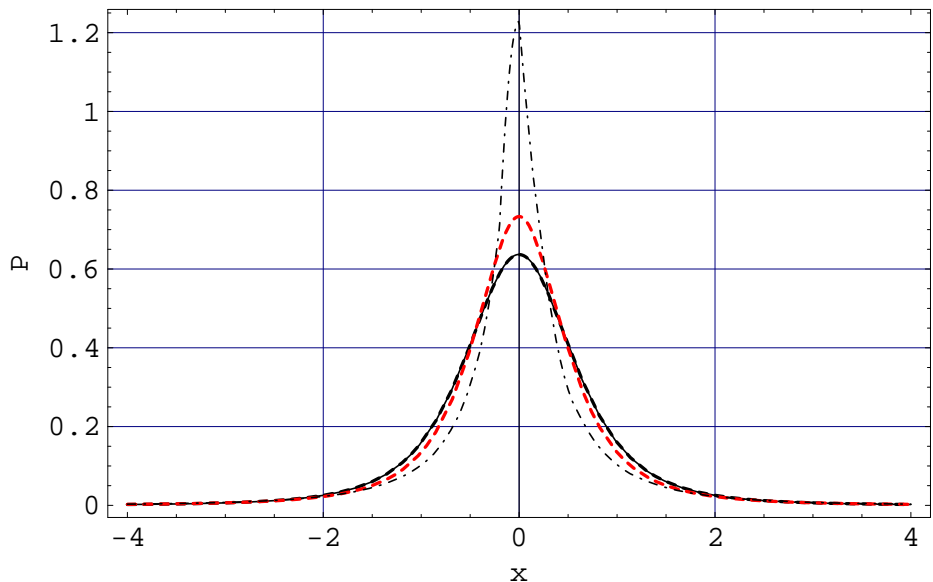}
\end{center}
\caption{ \label{Fig2} Lin-lin plot of fluid particle acceleration
probability density functions. Same notation as in
Figure~\ref{Fig1}.}
\end{figure}
%%%%%%%%%%%%%%%%%%%%%%%%%%%%%%%%%%%%%%%%%%%%%%%%%%%%%%%

Actually, one can represent the fluctuating energy dissipation
rate as $\epsilon = \bar\epsilon +\delta\epsilon$, with
$\langle\epsilon\rangle \equiv \bar\epsilon\not=0$ and zero mean
of the random fluctuating part, $\langle\delta\epsilon\rangle=0$.
In the experiments~\cite{Bodenschatz}, the measured mean energy
dissipation rate $\bar\epsilon$ varies by about 5 orders of
magnitude, from 9.01$\times$10$^{-5}$ to 9.16 m$^2\cdot$s$^{-3}$,
with $R_\lambda$ increasing from 140 to 970. Also, the variance of
$\delta\epsilon$ varies with approaching to the Kolmogorov scale.
Hence, both the effects should be taken into account.

We remind that the basic assumption of the model, $\beta \propto
\epsilon_r\tau$, used in reference~\cite{Beck} implies that $\beta
\propto u_1^2+u_2^2+u_3^2$ at Kolmogorov scale so that $\beta$ is
evidently $\chi^2$ distributed since the components of velocity
fluctuations, $u_i$, are known to be approximately normally
distributed with zero mean in accord to the
experiments~\cite{Bodenschatz}. This consideration is partially
justified from turbulence dynamics as the acceleration is
generally associated to small scales of the flow. The parameter
$q$ entering equations~(\ref{2}) and (\ref{4}) acquires the value
$q=(n+3)/(n+1)=3/2$, for the case of a linear drift force, since
there are $n=3$ independent random variables $u_i$~\cite{Beck}. It
should be emphasized that here $\beta$ is responsible only for a
fluctuating part of the energy dissipation.

Immediate testable consequences of the present model are the
probability density function $P(a)$ given by equation~(\ref{4})
and the contribution to the fourth order moment, $a^4P(a)$. The
recent experimental data~\cite{Bodenschatz2} established a
finiteness of the fourth order moment of the component of
acceleration, $\langle a^4\rangle$.

Plots of the distributions (\ref{1})-(\ref{3}), (\ref{4}), and the
resulting $a^4P(a)$ normalized to unit variance are shown in
Figures~1 and 2. The numerical value $a_c=39.0$ is obtained by a
direct fitting of the probability density function (\ref{4}) to
the experimental data~\cite{Bodenschatz2}. It is due to well
overlapping with the data points $P(a)$. Sensitivity of the
fitting is of an exponential type.

As one can see from the top panel of Figure~\ref{Fig1}, the
log-normal model (\ref{3}) (dot-dashed line) provides a good fit
to the experimental curve (thick dashed line) in the entire range
while the chi-square model (\ref{2}) (dashed line) and the
chi-square Gaussian model (\ref{4}) (solid line) are in a good
qualitative agreement with it except for that the chi-square model
implies increasing deviation at big $|a|$. From this point of
view, introducing of some truncation of the power law tails and
associated additional parameter is a necessary step to provide
consistency of the chi-square model with the experiment.

However, the central part of the distributions shown in
Figure~\ref{Fig2} reveals greater inaccuracy of the log-normal
model ($P(0)\simeq 1.23$) as compared with that of both the
chi-square and chi-square Gaussian models ($P(0)\simeq 0.65$)
which are almost not distinguishable in the region $|a|/\langle
a^2\rangle^{1/2}\leq 4$; see also recent work by Gotoh and
Kraichnan~\cite{Kraichnan0305040}. This is the main failing of the
log-normal model (\ref{3}) for $s^2=3.0$ although the predicted
distribution follows the measured low probability tails, which are
related to the Lagrangian intermittency, to a good accuracy. The
central region of the experimental curve (\ref{1}) ($P(0)\simeq
0.73$) contains most weight of the experimental distribution and
is the most accurate part of it, with the relative uncertainty of
about 3\% for $|a|/\langle a^2\rangle^{1/2}<10$ and more than 40\%
for $|a|/\langle a^2\rangle^{1/2}>40$~\cite{Mordant0303003}.

It should be noted that both the distributions (\ref{3}) and
(\ref{4}) show good agreement of $a^4P(a)$ (the contribution to
the kurtosis summarizing the peakedness of distribution) for small
and big $|a|$ but deviate from the experimental data for
intermediate $|a|$. This can be readily seen from the heights of
the peaks and from their positions in the bottom panel of
figure~1, $a/\langle a^2\rangle^{1/2}\simeq\pm 8.0$ and $\pm 7.1$
for maximal values of $a^4P(a)$ derived from equations~(\ref{3})
and (\ref{4}), respectively, at the given values of the
parameters, as compared to $\pm 10.2$ for the experimental curve
(\ref{1}). Also, in contrast to the chi-square model (\ref{2}),
the chi-square Gaussian model (\ref{4}) of the present work yields
a correct (decaying) behavior of the contribution to kurtosis at
big $|a|$ owing to the Gaussian truncation.

%% 4
\section{Conclusions}
\label{Sec:Conclusions}

(i) We have proposed a simple natural extension of the chi-square
distribution based stochastic model of fluid particle acceleration
in the developed turbulent flow which is found to be characterized
by the Gaussian truncated power-law type stationary probability
density function of the component of Lagrangian acceleration.
Despite one more parameter, $a_c$, has been invoked, such a
truncation and associated introducing of a new parameter is viewed
as a {\em necessary} step to provide consistency of implications
of the chi-square type model with the observed shape of
low-probability tails, which essentially characterize Lagrangian
intermittency.

(ii) We have made a comparison of the resulting distribution with
that of the chi-square model and the recently proposed log-normal
distribution based model. All these models fit the experimental
curve $P(a)$ for the transverse component of acceleration to a
more or less accuracy (the top panel of Figure~\ref{Fig1}).
However, the contribution to the fourth order moment of
acceleration was found to qualitatively discriminate among the
models (the bottom panel of Figure~\ref{Fig1}): the chi-square
model gives a qualitatively unsatisfactory behavior of the tails
of predicted contribution to kurtosis, $a^4P(a)$, while the
chi-square Gaussian model provides a good fit only for small
($|a|/\langle a^2\rangle^{1/2}<4$) and big ($|a|/\langle
a^2\rangle^{1/2}>30$) acceleration magnitude values. We found that
the Gaussian truncated power law tails are consistent with the
observed asymptotics of the acceleration probability density
function up to $|a|/\langle a^2\rangle^{1/2}=60$. The log-normal
model with $s^2=3$ meets the experimental data on acceleration
distribution in the entire range of measured accelerations but
implies a drastically bigger departure from the experimental data
in the central region (Figure~\ref{Fig2}), $|a|/\langle
a^2\rangle^{1/2}\leq 4$, which represents the most accurate part
and the most weight of the measured distribution. Also, both the
log-normal and the chi-square Gaussian model do not meet heights
and positions of the peaks of the experimental curve $a^4P(a)$.

(iii) The considered one-dimensional toy models of the Lagrangian
particle acceleration represent explicit examples of
parametrization that could be used as prototypical models.
Clearly, the choice of relevant forms of the distribution of the
parameter $\beta$ is phenomenological but this approach allows one
to deal with a part of statistics of complex temporal behavior of
fluid particle acceleration in the developed turbulent flow.

The problem of selecting of the underlying distribution of the
parameter $\beta$ has been addressed in the recent
work~\cite{Aringazin4}, in which we related this class of models
to the Heisenberg-Yaglom picture of the developed turbulence and
constructed the framework which incorporates the log-normal and
chi-square Gaussian models in a unified way as particular cases,
with the dependence $\beta(u)$ required to be a (monotonic) Borel
function of the stochastic Lagrangian velocity fluctuations $u$.

(iv) It should be noted that the gamma (chi-square) distribution
of the inverse temperature parameter has been considered from
various aspects~\cite{Beck,Johal,Aringazin}, is known to be in
correspondence with a nonlinear Langevin equation for fluctuating
temperature appearing in the Landau-Lifschitz theory of fluid
fluctuations~\cite{Bashkirov}, and shown to be of a universal
character for small fluctuations~\cite{Beck3}. However, due to the
analysis made in the present paper some {different} distributions
of intensive fluctuating parameters~\cite{Aringazin,Beck3,Beck4}
appear to be relevant to describe the experimental statistics of
fluid particle acceleration in the developed turbulence. In the
absence of a direct support from the Navier-Stokes equation, the
found departures from the experimental data can also be due to
shortcomings of the considered simple class of one-dimensional
stochastic models. Thus, it is appropriate to use a more general
stochastic approach, such as that with an inclusion of
multiplicative noise and nonlinear terms~\cite{Laval}, that
evidently would require several free parameters to be used.

We conclude by a few remarks.

We note that as it is common to Langevin statistical models of
developed turbulence, the considered class of models differs from
the phenomenological dimensional analysis (scaling relations)
approach since the former is based on a stochastic dynamical
framework dealing with time evolution of the Lagrangian
acceleration.

Since the introduced parameter $a_c$ at fixed $\beta_0$ measures
variance of the acceleration, equation~(\ref{6}), and the
Kolmogorov constant $a_0 \propto R_{\lambda}^{0.14}$ for high
$R_\lambda$ \cite{Bodenschatz}, the former may depend on the
Reynolds number. This dependence could be used to explain the fact
that the experimental $P(a)$ was found to have Reynolds number
dependent stretched exponential tails~\cite{Bodenschatz}.\\

\end{document}